\documentclass[12pt,preprint]{aastex}
\usepackage{natbib}

\def\wig#1{\mathrel{\hbox{\hbox to 0pt{%
          \lower.5ex\hbox{$\sim$}\hss}\raise.4ex\hbox{$#1$}}}}

\def\teff{T_{\rm eff}}
\def\fsed{f_{\rm sed}}

\shorttitle{Cloud Holes}
\shortauthors{Marley et al.}
\begin{document}

\title{A Patchy Cloud Model for the L to T Dwarf Transition}

\author{Mark S. Marley}
\affil{NASA Ames Research Center, MS-245-3, Moffett Field, CA 94035,  
U.S.A.; Mark.S.Marley@NASA.gov}

\author{Didier Saumon}
\affil{Los Alamos National Laboratory, Mail Stop F663, Los Alamos NM 87545; dsaumon@lanl.gov}

\and

\author{Colin Goldblatt}
\affil{Astronomy Department \& NASA Astrobiology Institute Virtual Planetary Laboratory, University of Washington, Box 351580, Seattle WA 98195, USA; cgoldbla@uw.edu}

\begin{abstract}

One mechanism suggested for the L to T dwarf spectral type transition is the appearance of relatively cloud-free regions across the disk of brown dwarfs as they cool.  The existence of partly cloudy regions has been supported by evidence for variability in dwarfs in the late L to early T spectral range, but no self-consistent atmosphere models of such partly cloudy objects have yet been constructed.  Here we present a new approach for consistently modeling partly cloudy brown dwarfs and giant planets.  We find that even a small fraction of cloud holes dramatically alter the atmospheric thermal profile, spectra, and photometric colors of a given object.  With decreasing cloudiness objects briskly become bluer in $J-K$ and brighten in $J$ band, as is observed at the L/T transition.  Model spectra of partly cloudy objects are similar to our models with globally homogenous, but thinner, clouds.  Hence spectra alone may not be sufficient to distinguish partial cloudiness although variability and polarization measurements are potential observational signatures. 
Finally we note that partial cloud cover may be an alternative explanation for the blue L dwarfs.

\end{abstract}


\section{INTRODUCTION}

As brown dwarfs cool over time, their atmospheres undergo a sequence of chemical and physical changes that result in an evolving emergent spectrum 
and--consequently--varying spectral types.  The most remarkable changes take place at the L to T type transition.   The latest L dwarfs have red near-infrared colors, strong CO absorption in $K$ band and relatively shallow water absorption bands modulating their spectra.
Over a small effective temperature range of only 100 to 
200$\,$K the spectrum rapidly changes to exhibit blue near-infrared color, weakening CO absorption, strengthening $\rm CH_4$ absorption, and deeper water bands (see \citet{Kir05} for a review).

The proximal cause of these changes is the loss of cloud opacity.  As clouds dissipate the visible atmosphere cools, bringing on the chemical change from CH$_4$ to CO.  Without clouds providing a significant, nearly gray opacity, flux can
emerge through molecular opacity windows in the J and H bands with
a brightening leading to a blueward color shift (Dahn et al. 2010, Tinney et al. 2003, Vrba et al. 2004). 
Studies of $\rm L + T$ binary dwarfs (e.g., Liu et al. 2006; Looper et al. 2008; Stumpf et al. 2010 and references therein) show that this brightening is an intrinsic signature of the transition and not the manifestation of some other effect in color-magnitude diagrams  \citep{Stu10}.

Two main underlying  causes of this loss in cloud opacity have been suggested.  In one view the atmospheric dynamical state changes, resulting in larger particle sizes that more rapidly `rain out' of the atmosphere, leading to a sudden clearing or collapse of the cloud \citep{Kna04, Tsu03, Tsu04}.  This view is supported by fits of spectra to model spectra \citep{Sau08}  computed with the \citet{Ack01} cloud model.  In that formalism a tunable parameter, $f_{\rm sed}$ controls cloud particle sizes and optical depth.  Larger $f_{\rm sed}$ yields larger particles along with physically and optically thinner clouds.  \citet{Cus08} and \citet{Ste09} have demonstrated that progressively later T dwarfs can be fit by increasing $f_{\rm sed}$ across the transition.  
Late T dwarfs are generally best fit by models that neglect cloud opacity.  \citet{Bur06} also suggest changes in cloud particle sizes as a possible mechanism.

The second view is inspired by thermal infrared images of the atmospheres of Jupiter and Saturn at $\sim 5\,\mu$m 
(e.g, Westphal 1969; Westphal et al. 1974; Orton et al. 1996; Baines et al. 2005).  In this spectral region gaseous opacity is very low, allowing flux from deeply seated, warm atmospheric regions to escape.  Higher lying clouds, however, locally reduce the emergent flux.  
As a result these planets take on a  mottled appearance with bright high-flux, low cloud opacity regions lying adjacent to cooler, darker, and cloudier regions (optical depth varies from $\sim 0$ to $\sim 20$ between these regions \citep{Ban98}).  \citet{Ack01} suggested that the arrival 
of such cloud holes near the end of the L spectral sequence may be responsible for the L to T transition.   \citet{Bur02} tested this hypothesis with a simple `toy model' by summing weighted contributions of the spectra of cloudy and cloudless models.  They showed that the observed 
$J$ band brightening across the transition could arise from decreasing cloud coverage.  Further support for this hypothesis comes from observations of L and T dwarf variability.  While previous studies were somewhat equivocal (summarized in \citet{Art09}), two early T dwarfs have recently been shown to 
have large near-infrared photometric variability \citep{Art09,Rad10} consistent with surface variations in cloud coverage modulated by rotation.

The approach to modeling holes of Burgasser et al.\ was highly simplistic.  The principal shortcoming being 
that it is not physically correct to combine the contributions of separate cloudy and cloudless models.  Deep in the atmosphere of a brown dwarf the 
entropy in the convection zone must match that of the interior. Thus the temperature at a given, deep, pressure level is expected to be horizontally constant.  
However for a fixed $\teff$ a cloudy atmosphere is everywhere hotter than a cloudless atmosphere.  As an 
example Figure 1 presents model atmosphere profiles for a uniform cloudy and a cloudless atmosphere following the techniques 
of \citet{Mar02} and \citet{Sau08}.  At depth, the cloudy profile is warmer than 
the cloudless profile by over 400$\,$K; the difference in some models is even greater.  Thus standard cloudy 
and  cloudless models cannot simultaneously be valid descriptions of the real atmosphere at two locations 
 even though both models are descriptions of an 
atmosphere with the same  $\teff$.  Clearly a new technique for 
self-consistently treating partly cloudy atmospheres is required.  While we do not yet understand why cloud holes might appear, 
we here present a new approach inspired from models of  Earth's
atmosphere to model their influence and apply our results to model the spectra and colors of L and T dwarfs.  

\section{Modeling Partly Cloudy Skies}

Instead of combining {\em separately} computed profiles for purely cloudy and cloud-free dwarfs we wish to construct a {\em single, global} temperature-pressure profile $T(P)$ that incorporates simultaneously the influences of both cloudy and 
cloud-free regions on the energy balance of the atmosphere.  The final profile should conserve the total flux while 
allowing for nearby atmospheric regions to have differing cloud -- but not thermal -- profiles.  
This conceptually allows clouds to be displaced by winds, updrafts, or downdrafts and change location, as long as the global mean cloud fraction is constant.

In three-dimensional terrestrial numerical weather 
prediction and climate models, clouds are typically smaller than the adopted computational grid scale. 
Various methods are used to treat this in the radiative transfer calculations, one of which is to consider separate cloudy and cloud-free \textit{sub-columns} in the \textit{same} model column, with a single $T(P)$ structure. This same approach can be used in a single column model of the Earth, allowing the global annual mean energy budget to be reproduced (Goldblatt \& Zahnle 2010).

Physically, such considerations are most important on Earth in the tropics. Taking the zonal mean $T(P)$ and moisture profile would give a local runaway greenhouse due to the high water vapor content. However, dry areas in the tropics caused by subsiding air act as ``radiator fins'', allowing radiation from the surface to escape to space (Peirrehumbert 1995). In brown dwarfs atmospheres, cloud free areas would be analogous to the subsidence regions in Earth's tropics.

To implement this approach we set the parameter $0\le h\le1$ to be the fraction of the atmosphere described by the cloud free sub-column, from which the local flux at some level in the atmosphere is $\cal F_{\rm hole}$. The remaining fraction $(1-h)$ is the cloudy sub-column, with local flux $\cal F_{\rm cloud}$. Both sub-columns share the same $T(P(z))$ profile but have different emergent flux, in general ${\cal F_{\rm hole}}>{\cal F}_{\rm cloud}$. We compute the total flux, ${\cal F_{\rm tot}}(z)$ through the atmosphere, which is used in the radiative-convective equilibrium calculation, as
$${\cal F_{\rm tot}}(z)=h{\cal F}_{\rm hole}(z) + (1-h){\cal F}_{\rm cloud}(z). \eqno(1)$$
We stress that this is not a combination of separate models, but rather conceptually represents two adjacent sub-columns in the atmosphere with the
same thermal profile and differing opacity.  
Our radiative-convective equilibrium model is then employed to solve for a single $T(P)$ atmospheric thermal profile (McKay et al. 1989; Marley \& McKay 1999) that carries net flux $\sigma T_{\rm eff}^4 = {\cal F_{\rm tot}}(z)$ through all
statically stable layers of the atmosphere.  
Layers that would be convectively unstable in pure radiative-equilibrium are iteratively adjusted to follow an adiabat.  The cloud profile and atmospheric chemistry are updated along with the atmospheric profile as it is converged to a solution, so the final thermal, chemical, and cloud profiles are all mutually self-consistent.  

A sampling of resulting model profiles is shown in Figure 1.  The hottest profile is for a global, homogeneous $f_{\rm sed}=2$ model while the coolest model assumes no cloud opacity.  Adding a small fraction of cloud free area to the $f_{\rm sed}=2$ cloudy 
model efficiently cools the atmospheric profile.  In the example shown setting $h=0.25$ produces a model that lies over one-third of the 
way between the fully cloudy ($f_{\rm sed}=2$) and the cloudless (nc) models in the region above 
$T(P)=T_{\rm eff}$.  Increasing the surface area fraction of holes to $h=0.5$  produces an even cooler model.  For comparison the figure also presents a homogeneous cloudy model with a much higher sedimentation efficiency, $f_{\rm sed}=4$.  
This model with globally low cloud opacity 
follows a $T(P)$ profile very similar to the partly cloudy model with thicker clouds ($f_{\rm sed}=2$) with 50\% of cloud-free 
regions.  We have found that this is generally the case: a partly cloudy model presents a similar 
thermal profile to a model with a thinner cloud, a point we return to in \S4.

\section{Photometry and Spectra of Partly Cloudy Dwarfs}

The synthetic spectrum for a converged $T(P)$ atmospheric profile is obtained from Equation (1). We compute absolute magnitudes and colors with
the ultracool dwarf evolution models of \citet{Sau08}.
Figure 2 shows synthetic near-infrared photometry from the partly cloudly models.  For a {\em fixed} $\teff$, 
as $h$ is increased the model colors briskly move to the blue in $J-K$ and $J-H$.  The $J$ band flux increases as $h$ increases from 0 to 0.75, but then 
dims slightly for cloud free models ($h=1$).  This is because atmospheres with even a small cloud cover are warmer in the atmospheric 
region from which the $J$ band flux emerges.   On the other hand, $M_H$ is relatively constant across the transition from $h=0$ to $h=1$ 
at constant $\teff$ but shows a similar dimming as $h \rightarrow 1$. 

The evolution of model dwarfs with fixed $f_{\rm sed}$ produces trajectories that do not exhibit a rapid L to T 
transition as a global homogenous cloud sinks too gradually below the photosphere (Burrows et al. 2006; Saumon \& Marley 2008, Fig. 4).  With the \citet{Ack01} cloud
model, the L to T dwarf transition can only be modeled with an increase of the cloud sedimentation parameter $f_{\rm sed}$.  Figure 2 
demonstrates that the transition can also be described as a progressive increase in cloud-free areas at a fixed $f_{\rm sed}$ and
$\teff \sim 1200\,\rm K$.  The L/T transition dwarf colors and the J band brightening are well fit by this approach.  However, 1200$\,$K is slightly 
cooler than the observed $\teff$ temperature of the transition of $\sim 1300\,\rm K$ (Golimowski et al. 2004; Stephens et al. 2009) and if $\teff$ falls appreciably across the transition the J band bump would be weakened (Fig. 2).
For a different choice of $f_{\rm sed}$ and model gravity a different transition $T_{\rm eff}$ would be expected. 

As noted in Saumon \& Marley, there is an offset of $\sim 0.3$ to the blue in the $J-K$ color of models from the bulk 
of L and T dwarf photometry.  This may arise from shortcomings in the $K$ band pressure-induced opacity 
of molecular hydrogen.  Nonetheless, the trend of the late T dwarf $J-K$ color can be better 
reproduced with models with $h \sim 0.5-0.75$.  In the $M_H$ vs. $J-H$ diagram, the behavior of field L dwarfs and
late T dwarfs are better reproduced by the cloudy $f_{\rm sed}=2$ and the cloudless sequences, respectively although the 
latter would also be better matched with partly cloudy models with $h \sim 0.5 - 0.75$.  This suggests the spectra of late T dwarfs could be influenced by clouds, contrary to the usual assumption \citep{Bur10}.  

\section{Distinguishing Partly Cloudy Dwarfs}

In Saumon \& Marley (2008) we demonstrated that by increasing the cloud sedimentation efficiency $f_{\rm sed}$ as a brown dwarf cools
from $T_{\rm eff}=1400$ to 1200$\,$K, 
the predicted model colors reproduce those across the L to T transition.  In the previous section we likewise showed that 
increasing fractional cloudiness -- at fixed $f_{\rm sed}$ -- has the same result.  This leads us to consider how to distinguish the two cases.

As shown in Figure 1 a partly cloudy $T(P)$ profile (based on a $f_{\rm sed}=2$ cloud) can be nearly identical 
to a model with a thinner homogeneous cloud but the spectrum from these two models are not necessarily 
the same because the former uses Eq.\ (1) to compute the flux.
In the partly cloudy case some flux from deep, hot regions of the atmosphere (${\cal F}_{\rm hole}$ is escaping through the clear regions
that are otherwise totally shielded by the cloud in the homogeneous case (see the middle panel of Figure 7 of \citet{Ack01}).  
Thus we expect that even for identical $T(P)$ profiles the emission spectra will differ.

Indeed that is the case as shown in Figure 4 which shows spectra computed from profiles shown in Figure 1.  Focusing on the $J$ band, which features the lowest molecular opacity and the deepest atmospheric window in the near-infrared 
\citep{Ack01}, the greatest flux is found for the cloudless model.  The homogenous cloudy $f_{\rm sed}=2$ model is 
faintest, with the $f_{\rm sed}=4$ case falling in between.  Even though the partly cloudy $h=0.5$ model has essentially the 
same thermal profile as the $f_{\rm sed}=4$ model, it is brighter in the $J$ band because some flux is escaping from deeper in the 
atmosphere.  Of course flux conservation requires that the partly cloudy model must be fainter at other wavelengths, 
here in $K$ band.  Thus the model spectra are increasingly bluer in the near-infrared from $\fsed=2$ to $\fsed=4$ to the partly cloudy model.
Figure 4 shows that the partly cloudy model is very close to the $\fsed=4$ homogeneous cloudy model at wavelengths where the flux
is low and emitted from the upper atmosphere, and intermediate between $\fsed=4$ and cloudless in the $JHK$ flux peaks where
part of the flux comes from the deeper atmosphere.  Thus, for a given $\teff$ and gravity, the presence of holes in the cloud cover
have a discernible effect on the near-infrared spectrum.

For a given observed spectrum, where $\teff$, gravity and composition are not known {\it a priori}, how can we distinguish
a partly from a homogeneous cloudy atmosphere?  We explored this problem by fitting the near-infrared partly cloudy model spectra
with our large library of cloudy models, using the method of  \citet{Cus08}, and allowing $\teff$, gravity, and $\fsed$
to vary freely.  We find that in general,
the best fitting cloudy model has the same gravity, the same $\teff$ (or slightly higher by $\sim 100\,$K) and a higher $\fsed$
depending on  $h$.  The fitted cloudy spectra are close to the partly cloudy spectra, and
some of the differences can be attributed to the grid spacing of the cloudy models. It appears that the $JHK$ colors of
a partly cloudy model ($\teff$, $g$, $\fsed$, $h$) can be well matched with a cloudy model ($\teff ^\prime$, $g ^\prime$, $\fsed^\prime$).
The differences in the near-infrared spectrum are thus subtle
and comparable to the differences found between observed spectra and the best fitting models (e.g. \citet{Cus08,Ste09}).
Given that even with bolometric luminosity measurements $\teff$ temperatures cannot yet be measured to much 
better than $\sim 50\,\rm K$ precision, effective temperature cannot yet provide a strong constraint on $h$.
The most promising avenue to distinguish the two models in the L/T transition spectroscopically appears to be strong 
spectral features that are very sensitive to temperature
and that are formed in the deeper, hotter regions of the atmospheres, such as the 0.99$\,\mu$m band of FeH and the alkali doublets of
Na I ($1.14\,\mu$m) and K I ($1.175$ and $1.25\,\mu$m). Those should become stronger as the cloud coverage decreases. 

Given these considerations, variability may be the best method of characterizing cloud 
patchiness in brown dwarfs.  Partly cloudy dwarfs may be variable if the lengthscales of the cloudy 
patches are large enough and if the viewing inclination is not near pole on.  Variations in the geometry
of the cloud cover (``weather'') would also lead to detectable variations in brightness on time scales different from
the rotation period. Indeed there is 
indication that a sizable fraction of L dwarfs are variable \citep{Gel02, Koen03}.  Sample sizes are as yet too 
small to determine if blue L dwarfs or L/T transition objects show greater variability than other L dwarfs, 
but systematic studies would elucidate any trends.

Polarization is another possible way to distinguish partly from fully cloudy objects.  
Cloud free brown dwarfs should not show appreciable polarization \citep{Sen09}. \citet{Sen10} 
demonstrated that very rapidly rotating, low gravity dwarfs with homogenous clouds can show sizeable 
polarization, but only for rotation periods less than about 4 hours. However partly cloudy L dwarfs may 
show a polarization signal at $I$ band consistent with that observed in some objects \citep{Men04}.    
Measurable polarization in a slowly rotating or high gravity brown dwarf would 
point to the presence of inhomogeneous cloud cover. 

\section{Discussion and Conclusions}
There is growing evidence that the transition from the L to the T spectral class happens over a small effective temperature range \citep[e.g.,][]{Ste09}.  It is difficult for any model of a globally uniform, homogenous cloud to either sink out of sight or precipitate rapidly enough to account for the observed rapid change in spectral properties.  Instead the transition may arise as holes appear in an otherwise uniform global cloud deck.

We have presented the first self-consistent method for computing one-dimensional global atmospheric profiles 
appropriate to an atmosphere which has both clear and cloudy patches.  The model does not depend upon the 
physical sizes of the patches, nor does it explain why patches might appear, but rather  parameterizes the global fraction of clear and cloudy regions.    
The resulting $T(P)$ profiles converged to by our model are intermediate in temperature between hot, 
cloudy models and colder, cloudless ones.  We confirm the conclusion of \citet{Bur02}, who used a more 
simplistic model to argue that patchiness in brown dwarf clouds is a plausible mechanism for the L to T type transition.  

In addition to the L to T transition, patchy clouds have been suggested by \citet{Fol07} as a possible mechanism to explain those L dwarfs that have unusually blue near-infrared colors \citep{Cru03, Kna04, Bur08}.  \citet{Bur08} demonstrated that a $T_{\rm eff}=1700\,\rm K$ model with thin ($f_{\rm sed}=4$) clouds provides a good spectral match to the blue L dwarf 2MASS J1126--5003.  We find that in this effective temperature regime a model with thicker ($f_{\rm sed}=2$) and 50\% cloud coverage produces a comparable spectrum.  Thus we agree with \citet{Fol07} that partial cloudiness is potential alternative mechanism to thin global clouds for the blue L dwarfs.  Differences in cloud coverage might result from different rotational rates, gravities, or even viewing geometries (pole vs. equator).  

The growing evidence for temporal photometric variability at the L to T transition \citep{Art09,Rad10} supports the plausibility of partial cloudiness being responsible for the change in colors across the transition.  Future 3-dimensional simulations of brown dwarf atmospheres \citep[e.g.,][]{Fre10} coupled with a new generation of cloud models will likely help to elucidate the underlying mechanism responsible for cloud fragmentation at $\teff \sim 1200$--1300$\,$K.  In the meantime additional observations of brown dwarf photometric variability and observations of L/T binaries will shed light on this poorly understood phase of their evolution.

\section{Acknowledgements}

We thank M.C. Cushing for performing partly cloudy fits and the anonymous referee for helpful comments.  NASA provided support for this work via the
Planetary Atmospheres Program (MSM and CG), the Spitzer Space telescope Theoretical Research Program (DS), and Astrobiology Institute's Virtual Planetary Laboratory Lead Team (CG).

\clearpage
\begin{figure}
\includegraphics[angle=0.0,scale=0.8]{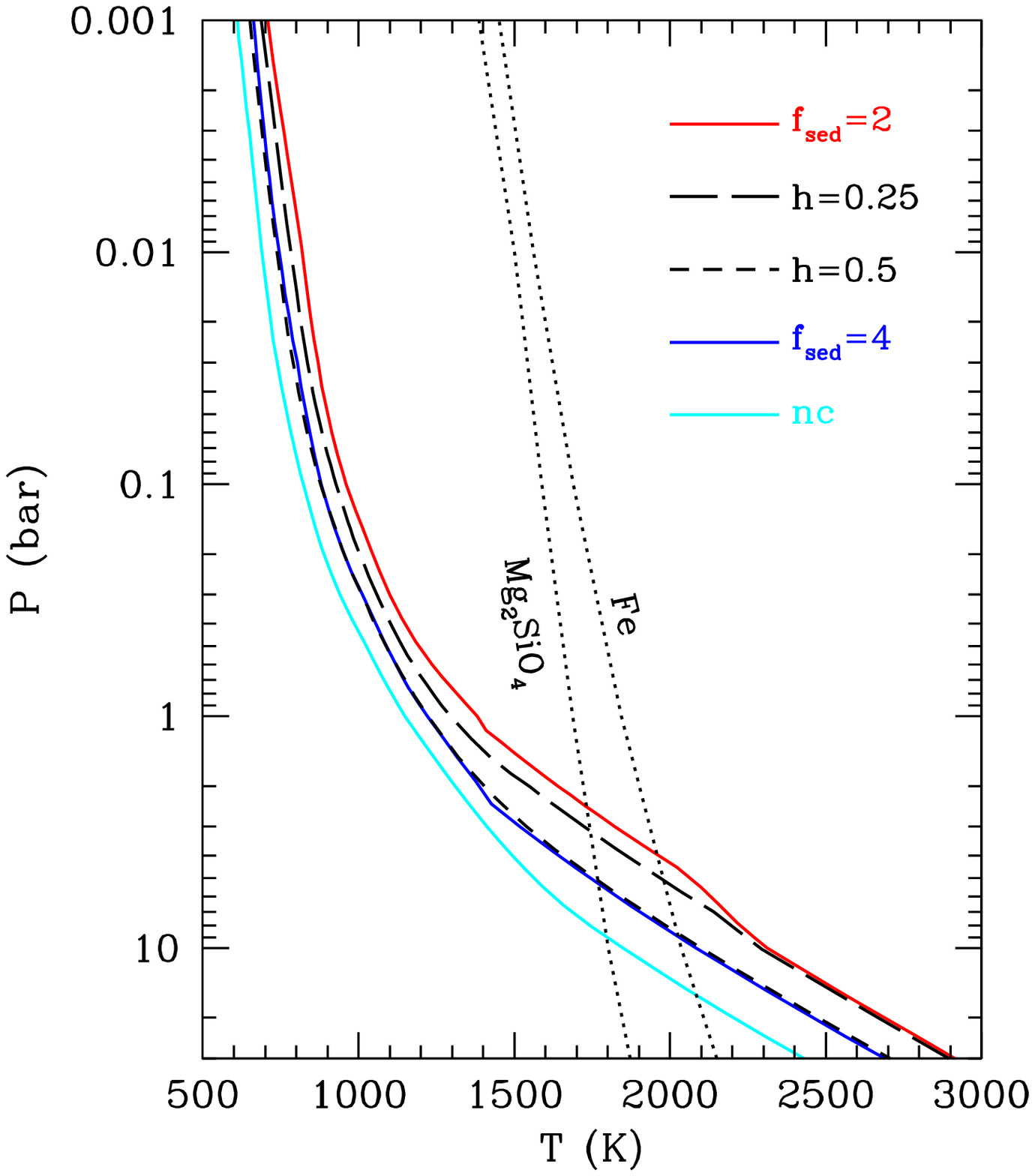}
\caption{Model atmosphere temperature-pressure profiles with $T_{\rm eff} = 1400\,\rm K$, $\log g {\rm (cm/s^2)} = 5$
and solar metallicity.  
Solid lines are for atmospheres with horizontally homogenous cloud cover (labeled $f_{\rm sed}=2$ and 
4) or no clouds (labeled nc).  Two partly cloudy models with hole fraction $h=0.25$ and 0.5 
and based on a $f_{\rm sed}=2$ cloud model are shown with dashed lines. The condensation curves of iron
and forsterite are shown with dotted lines.}

\label{fig1}
\end{figure}

\clearpage
\begin{figure}
\includegraphics[angle=0.0,scale=0.9]{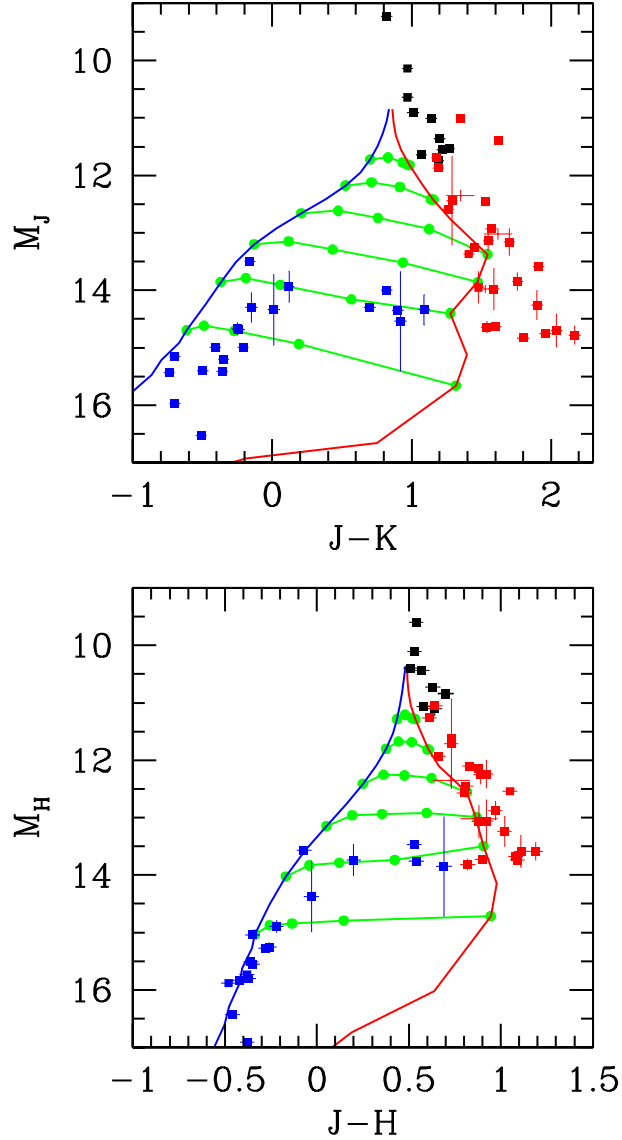}
\caption{Near-infrared model color-magnitude diagrams (MKO system).  Red and blue lines, respectively, 
show model colors for cloudy sequence with $f_{\rm sed} = 2$ and cloudless models (both $\log g = 5$).  Horizontal green lines 
connect partly cloudy model 
colors with $T_{\rm eff} = 1000$ to 2000 K in steps of 200K (from bottom to top).  Partly cloudy models (green dots) are for 
cloud-free fractions $h$ of 0, 0.25, 0.5, 0.75 and 1, from right to left.
Square symbols show field dwarfs with M dwarfs in black, L dwarfs in red and T dwarfs in blue. 
The photometry is primarily from \citet{leggett02} and \citet{Kna04}. Additional sources are given in
\citet{Sau08} (Fig. 7).}

\label{fig2}
\end{figure}

\clearpage
\begin{figure}
\includegraphics[angle=-90.0,scale=.60]{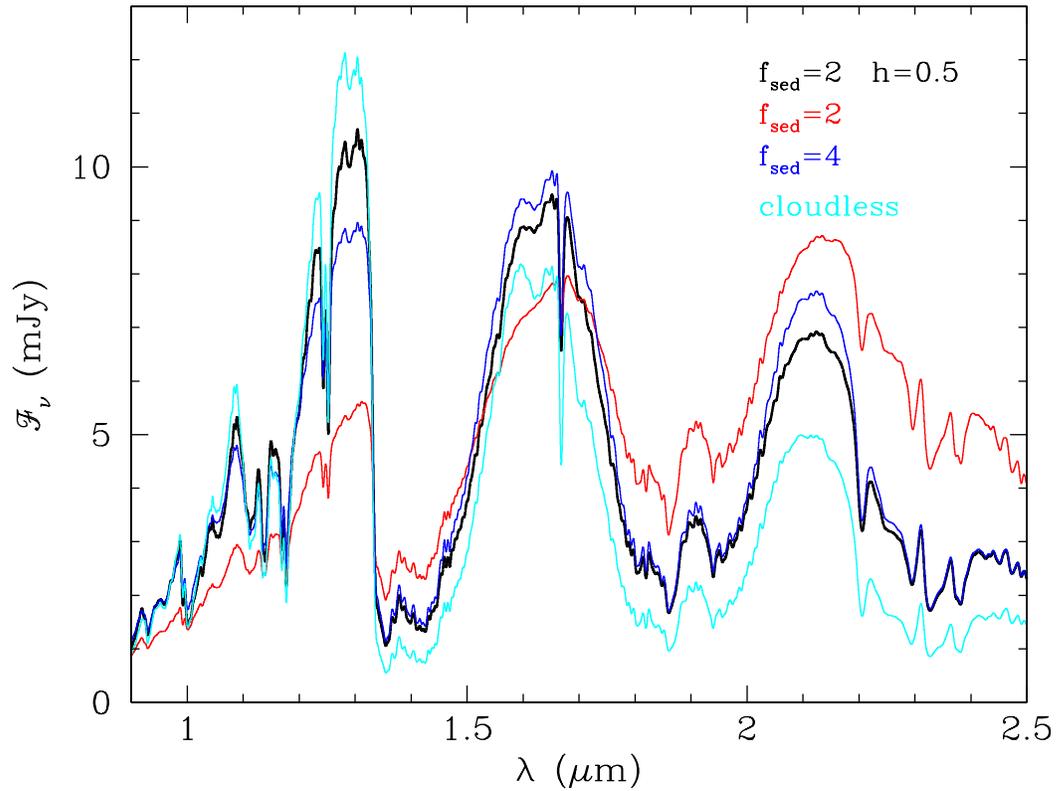}
\caption{Absolute model fluxes ($d=10\,$pc) for various cloud treatments, all with $\teff=1400\,$K, $\log g = 5$
and solar metallicity.  
Homogeneous cloudy spectra are shown for $f_{\rm sed}=2$ and 4 along with a cloudless spectrum.
The  partly cloudy model with $h=0.5$ uses a $f_{\rm sed}=2$ cloud.  
These spectra are computed from the thermal profiles shown in Figure 1.}

\label{fig3}
\end{figure}

\end{document}